\documentclass[prl, twocolumn,superscriptaddress,preprintnumbers,a4paper,amsmath,amssymb,showpacs,floatfix]{revtex4}
\usepackage{graphicx}
\usepackage{color}\color[rgb]{0.000,0.000,0.000} 
\usepackage{amssymb} 
\usepackage{amsmath} 
\usepackage{multirow}

\begin{document}
\newcommand{\mn}{MnS$_{2}$}
\newcommand{\te}{$_{t}$}
\newcommand{\ot}{$_{o}$}
\newcommand{\Tc}{T$_{C}$}
\newcommand{\Ts}{T$_{s}$}
\newcommand{\Tn}{$T_\mathrm{N}$}
\newcommand{\MuB}{$\mu_\mathrm{B}$}
\title{Spin-driven symmetry breaking in the frustrated \textit{fcc}~magnet \mn}

\author{Simon A. J. Kimber}
\email[Email of corresponding author:]{kimber@esrf.fr}
\affiliation{European Synchrotron Radiation Facility (ESRF), 6 rue Jules Horowitz, BP 220, 38043  Grenoble Cedex 9, France}

\author{Tapan Chatterji}
\affiliation{Institute Laue Langevin (ILL), 6 rue Jules Horowitz, BP 220, 38043  Grenoble Cedex 9, France.}

\date{\today}

\begin{abstract}
We report the characterisation of natural samples of the cubic pyrite mineral MnS$_{2}$  using very high resolution synchrotron X-ray diffraction techniques. At low temperatures we find a new low temperature polymorph, which results from coupling between magnetic and lattice degrees of freedom. Below the magnetic ordering temperature T$_{N}$= 48 K, we detect a pseudo-tetragonal distortion with a tiny \textit{\textbf{c}}/\textit{\textbf{a}} ratio of 1.0006. The structure can be refined in the space group $Pbca$~The symmetry lowering reduces magnetic frustration in the $fcc$ \ Mn$^{2+}$ \ lattice and is likely responsible for the previously reported lock-in of the magnetic propagation vector. This behaviour is similar to the frustration driven symmetry breaking reported in other three-dimensional Heisenberg magnets like the chromate spinels.\end{abstract}
\maketitle
\noindent{\large{\textbf{Introduction}}}\\
{\normalsize The archetypal pyrite material FeS$_{2}$ is the most commonly found sulfide mineral, and is popularly known as 'fools gold' due to its metallic lustre. Besides the obvious geological interest, pyrite structured materials are also notable for their electronic and magnetic properties. For example, CuS$_{2}$ shows superconductivity \cite{CuS2}~below $\sim$ 2.5 K and the itinerant ferromagnet CoS$_{2}$ can be tuned through a quantum critical point \cite{CoS2}~at high pressures and low temperatures.\\
The pyrite structure (Fig. 1) is made up of a rock salt arrangement of A$^{2+}$ cations and isolated X$^{2-}$ molecules. The latter are aligned along the [111] directions, as evidenced by the presence of weak X-ray and neutron reflections that break the $fcc$ systematic absence conditions. The symmetry of these materials is thus reduced to $Pa\bar{3}$.  Some materials, notably alkali metal superoxides like NaO$_{2}$, may be regarded as molecular crystals, and show an instability to a plastic rotator phase with average $Fm\bar{3}m$ symmetry near room temperature\cite{NaO2}. In contrast, the results of Raman scattering \cite{raman}~show that the transition metal pyrites exhibit continuous bonding, despite the quasi-molecular nature of the pyrite structure. \\
\begin{figure}[tb!]
\begin{center}
\includegraphics[scale=0.38]{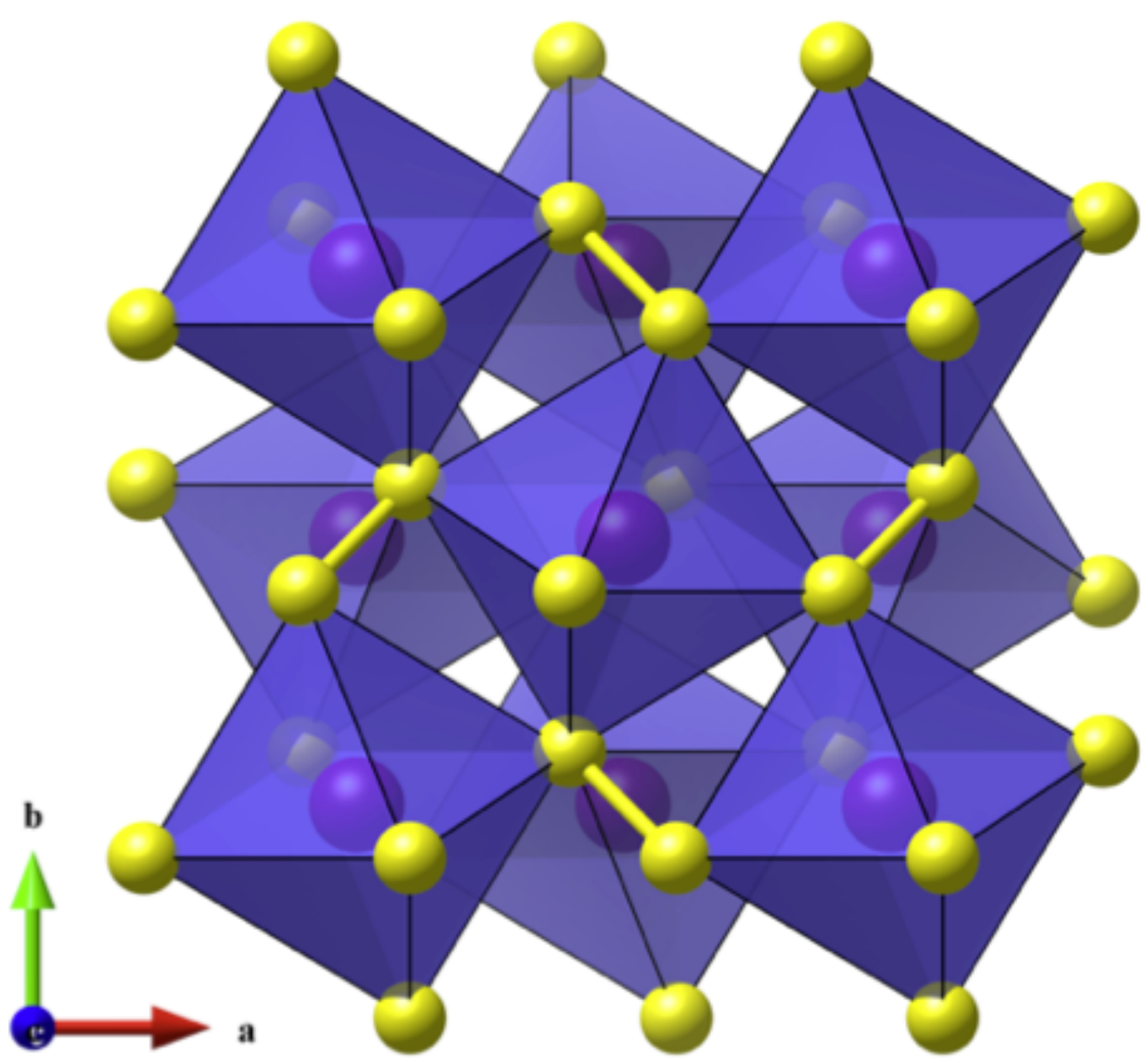}
\caption{(color online) a) Structure of $Pa\bar{3}$~\mn. The purple spheres represent the Mn$^{2+}$ cations, while the yellow dumbells show the S$_{2}^{2-}$ units. The octahedral coordination around the Mn$^{2+}$ cations is highlighted by the shaded coordination polyhedra.}
\label{Fig1}
\end{center}
\end{figure}
Here we report a re-examination of the structure of natural samples of Hauerite (MnS$_{2}$) as a function of temperature \cite{MnS2_TC1}.  This mineral is magnetic due to the unpaired spins associated with the Mn$^{2+}$ cations. To a first approximation, the crystal field in pyrites is purely octahedral. In combination with Hund's rule coupling, this results in a spin-only S=5/2 high spin state. The single-ion moments are thus isotropic, and the magnetism of \mn~should be described by a classical three-dimensional Heisenberg model. Previous neutron diffraction studies have shown \cite{Hastings}~that the Mn$^{2+}$ spins order at T$_{N}$ = 48.2 K. However, the exact magnetic structure remains unresolved, although polarized single crystal neutron studies show that it must be co-planar \cite{MnS2_TC2}, and is a type-III antiferromagnet. The magnetic transition is apparently first-order and accompanied by twinning. Curiously, the magnetic diffuse scattering above T$_{N}$ appears at an incommensurate wavevector, while the magnetic order has a commensurate \textbf{\textit{k}} = (1,1/2,0) propagation vector \cite{MnS2_TC3}.  The locking in of the wave vector to a commensurate position might indicate strong spin-lattice coupling. While little has been reported at ambient pressure, with no detected anomalies in e.g. thermal expansion \cite{MnS2_TC1}, we recently discovered \cite{PNAS}~  such coupling in MnS$_{2}$ under pressure. Above 11.7 GPa a spin-state transition is found, commensurate with a structural distortion to monoclinic symmetry. The magnetic moments are globally quenched by the formation of Mn-Mn dimers. Some evidence for spin-lattice coupling at low temperatures has also been previously reported. In 1996, Chattopadhyay and Liss performed pioneering hard x-ray triple axis experiments on \mn. By following the (10,0,0) reflection as a function of temperature, they observed \cite{KDL1}~a tiny splitting at the N\'eel temperature.  However, this result was never formally investigated \cite{KDL2}~and no detailed crystallographic measurements of sufficient resolution have subsequently been performed at low temperatures. Motivated by this work, we re-examined the ambient pressure structure at temperatures spanning T$_{N}$. By using very high resolution synchrotron X-ray diffraction, we show that the magnetic ordering is accompanied by a structural distortion to  pseudo-tetragonal symmetry. This presumably breaks magnetic frustration, allowing the establishment of long-range order. This observation is somewhat similar to the magneto-elastic coupling seen \cite{ZnCr2O41,ZnCr2O42}~in the \textit{A}Cr$_{2}$O$_{4}$ spinels, which are also frustrated three-dimensional Heisenberg systems.}\\
\\
\noindent{\large{\textbf{Experimental}}}\\
\begin{figure}[tb!]
\begin{center}
\includegraphics[scale=0.65]{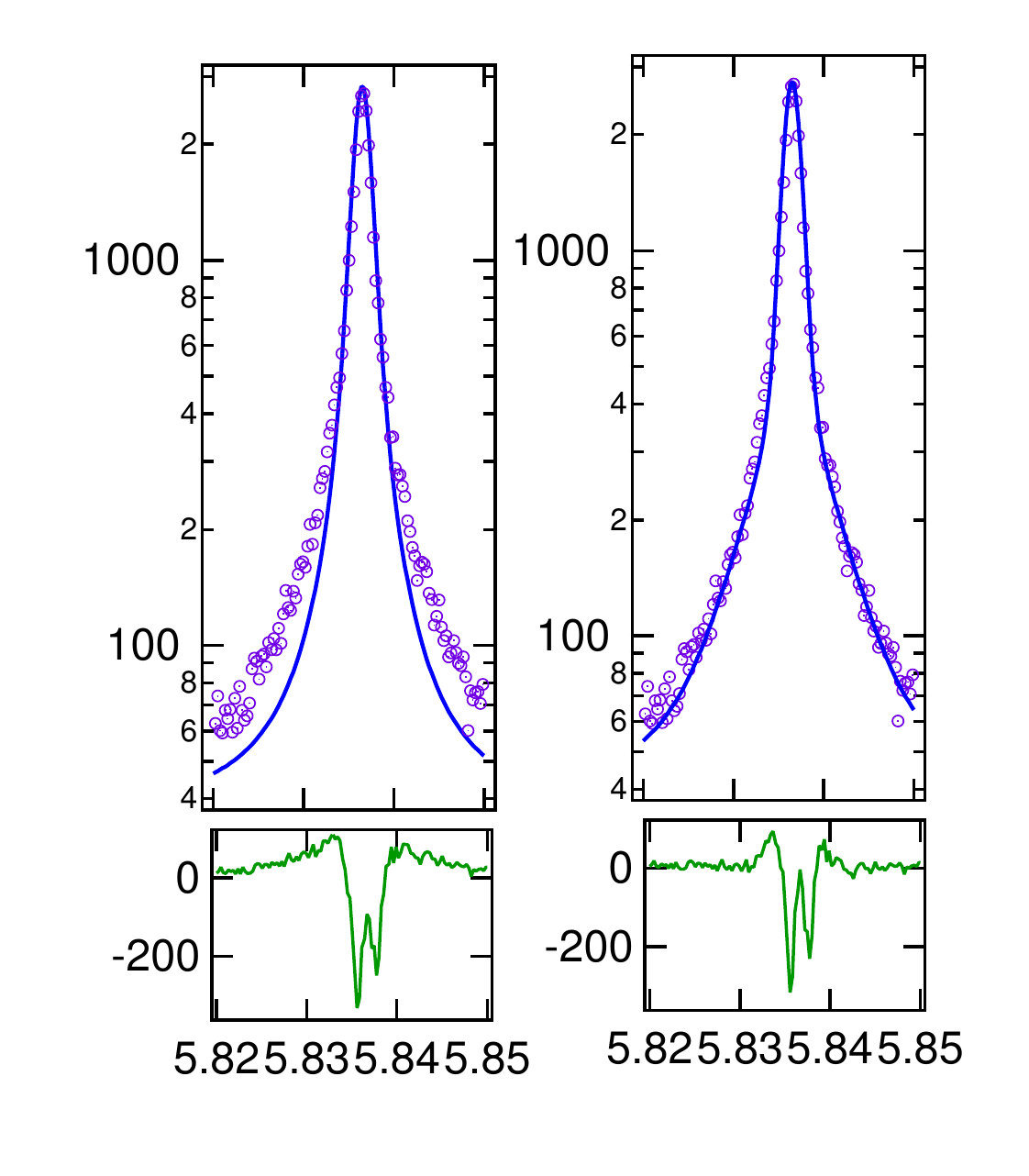}
\caption{(color online) (a) Observed, calculated and difference profiles for a portion of the Rietveld fit to the 55 K diffraction profile of \mn. Here a standard pseudo-voigt profile was used; b) Observed calculated and difference profiles for the Rietveld fit to the same data using a double Lorentzian peak shape.}
\label{Fig1}
\end{center}
\end{figure}
{\normalsize Natural single crystal samples of MnS$_{2}$ \ were obtained from the M\"unster Geologisches Museum. Note that these are the same samples used for earlier single crystal neutron and high pressure diffraction studies \cite{MnS2_TC1,MnS2_TC2,MnS2_TC3,PNAS}. We used a small part of a large crystal of octahedral habit. This was crushed into a fine powder for our diffraction measurements using an agate mortar and pestle. As reported previously, the elemental composition of our sample was checked using X-ray fluorescence measurements (see the supplementary info in Ref. \cite{PNAS}).\\
\begin{figure}[tb!]
\begin{center}
\includegraphics[scale=0.38]{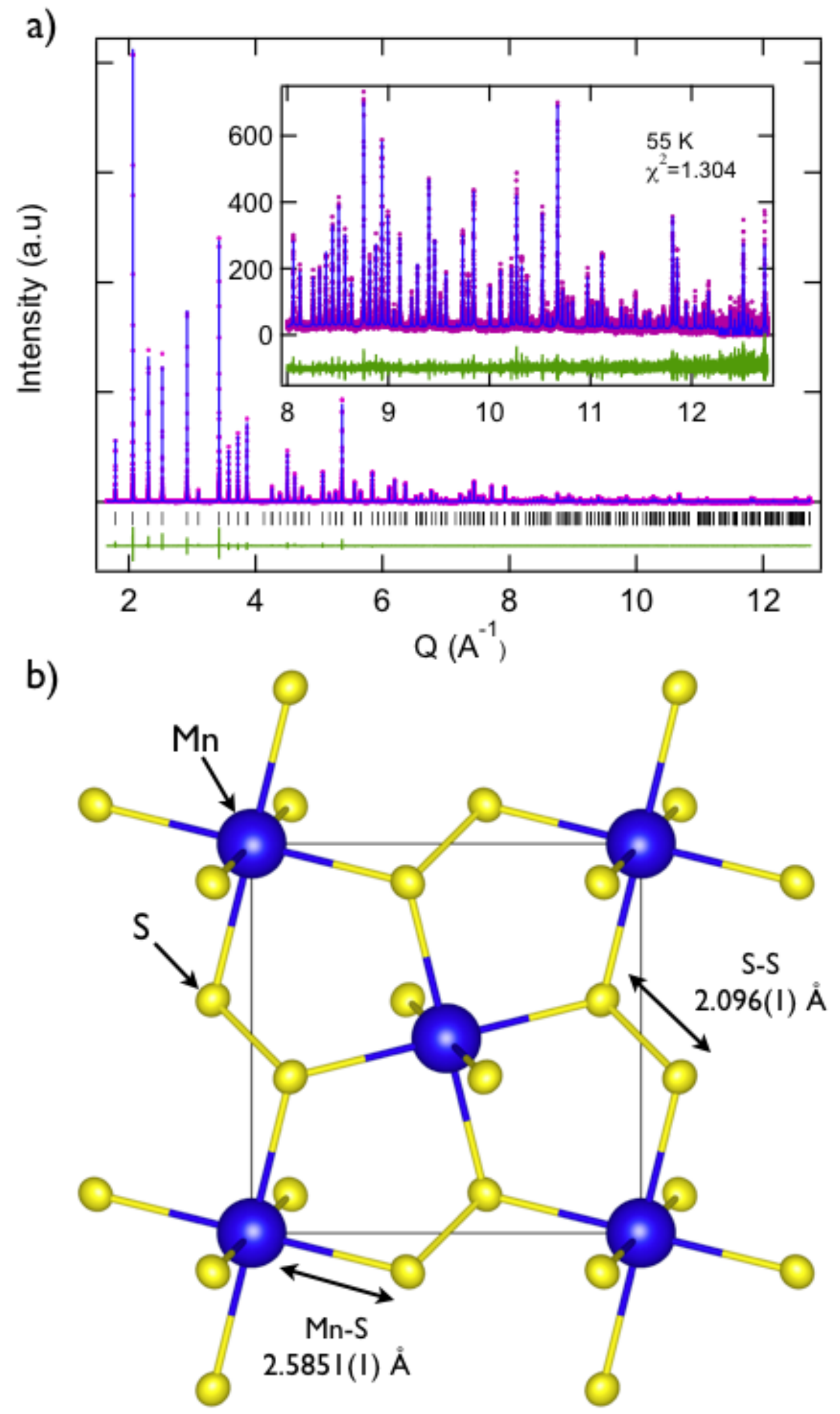}
\caption{(color online) (a) Observed, calculated and difference profiles for the  complete Rietveld fit to the 55 K diffraction profile of \mn. The high angle region is expanded in the inset to highlight the excellent fit over the entire Q-range measured; b) Structural details of the refined model, highlighting the S-S and Mn-S distances.}
\label{Fig1}
\end{center}
\end{figure}
High resolution powder diffraction patterns were collected on capillary samples at 10 and 55 K on the crystal analyser diffractometer ID31 at the ESRF, using an X-ray energy of 31 KeV. Note that at the time of writing, this instrument has moved to a new insertion device port at ESRF, ID22. For typical materials, the angular resolution of this instrument is usually dominated by sample broadening, with an instrumental contribution \cite{ID31}~of around 0.003 $^{\circ}$ FWHM. Due to the decrease in signal at higher angles caused by the x-ray form factor, a counting scheme which overemphasised the data at high momentum transfers was employed \cite{collect}. The total collection time was of the order of 3 hours for each data set. After appropriate normalisation, this resulted in data sets containing useful diffraction intensities over the range 1 $<$ Q $<$ 14.5 \AA$^{-1}$. More rapid data collections (ca. 2 min) were performed on cooling through the phase transition. The diffraction patterns were analysed using the Rietveld method as implemented in the GSAS program \cite{gsas}~with the EXPGUI interface \cite{expgui}.}\\

\noindent{\large{\textbf{Results}}}\\
{\normalsize
The diffraction profile of \mn~at 55 K could be indexed by the $Pa\bar{3}$~pyrite cell with \textbf{$a$} = 6.088821(4) \AA. However, during the initial stages of the structure refinement, we noticed that the standard peak shape functions included in GSAS failed to fit the tails present on each Bragg reflection. As shown in Fig. 2 (which uses a logarithmic scale), even Lorentzian functions failed to fit this feature. The initial cycles of refinement converged with $\chi^{2}\sim$5, showing rather poor agreement between the data and model calculation.  A similar observation was made for data collected on the spinel magnetite using the BM16 beam line at ESRF, when it previously housed the high resolution diffractometer. In that case, the sample was also prepared from a large high-quality single crystal \cite{jon}. The peak shape was attributed to a small 'tail' of crystallites generated by manual grinding. Taking inspiration from this early work, we obtained much better fits to the diffraction profiles using a double Lorentzian peak shape (Fig. 2). This approach was used for both the 55 and 10 K data sets. After accounting for this microstructural detail, an excellent Rietveld fit was obtained (Fig. 3a), which yielded the agreement factor $\chi^{2}$=1.304 and the results listed in Table I. The degree of precision obtained on the bond distances and fractional coordinates (e.g. the Mn-S distance of  2.5851(1) \AA~shown in Fig. 2b) highlights the excellent quality of data obtainable on ID31, as well as the very long-range structural coherence in \mn. We did not observe any significant impurity phases, bar a few small peaks with an intensity less than 1 \% of those from the main phase. Due to their scarcity, we were unable to identify these trace impurities. Together with the XRF results reported previously \cite{PNAS}, this indicates that our sample is of remarkable purity and stoichiometry for a natural mineral.\\
\begin{figure}[tb!]
\begin{center}
\includegraphics[scale=0.5]{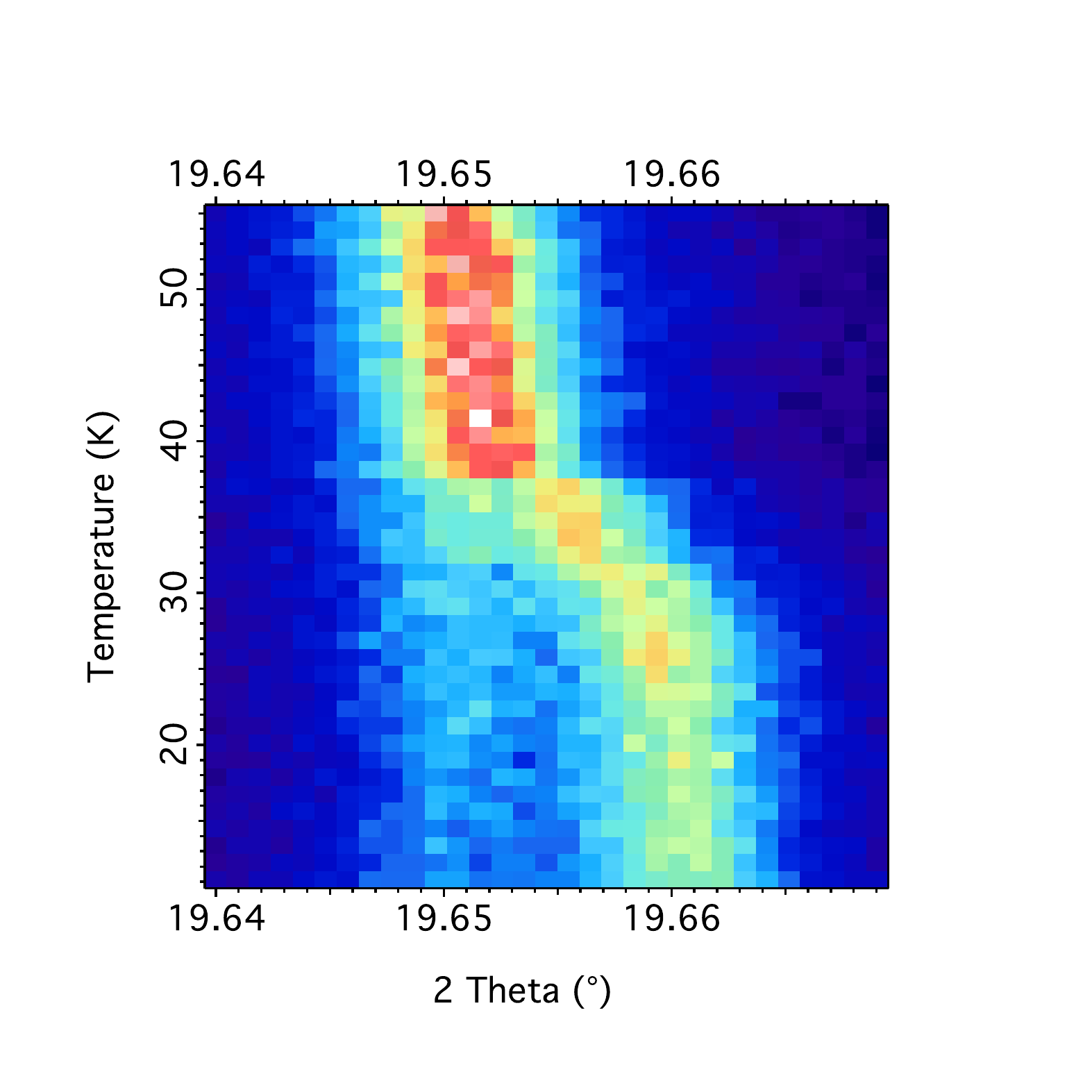}
\caption{(color online) The effect of temperature on the (511)$_{c}$ reflection}
\label{Fig1}
\end{center}
\end{figure}

\begin{figure}[tb!]
\begin{center}
\includegraphics[scale=0.45]{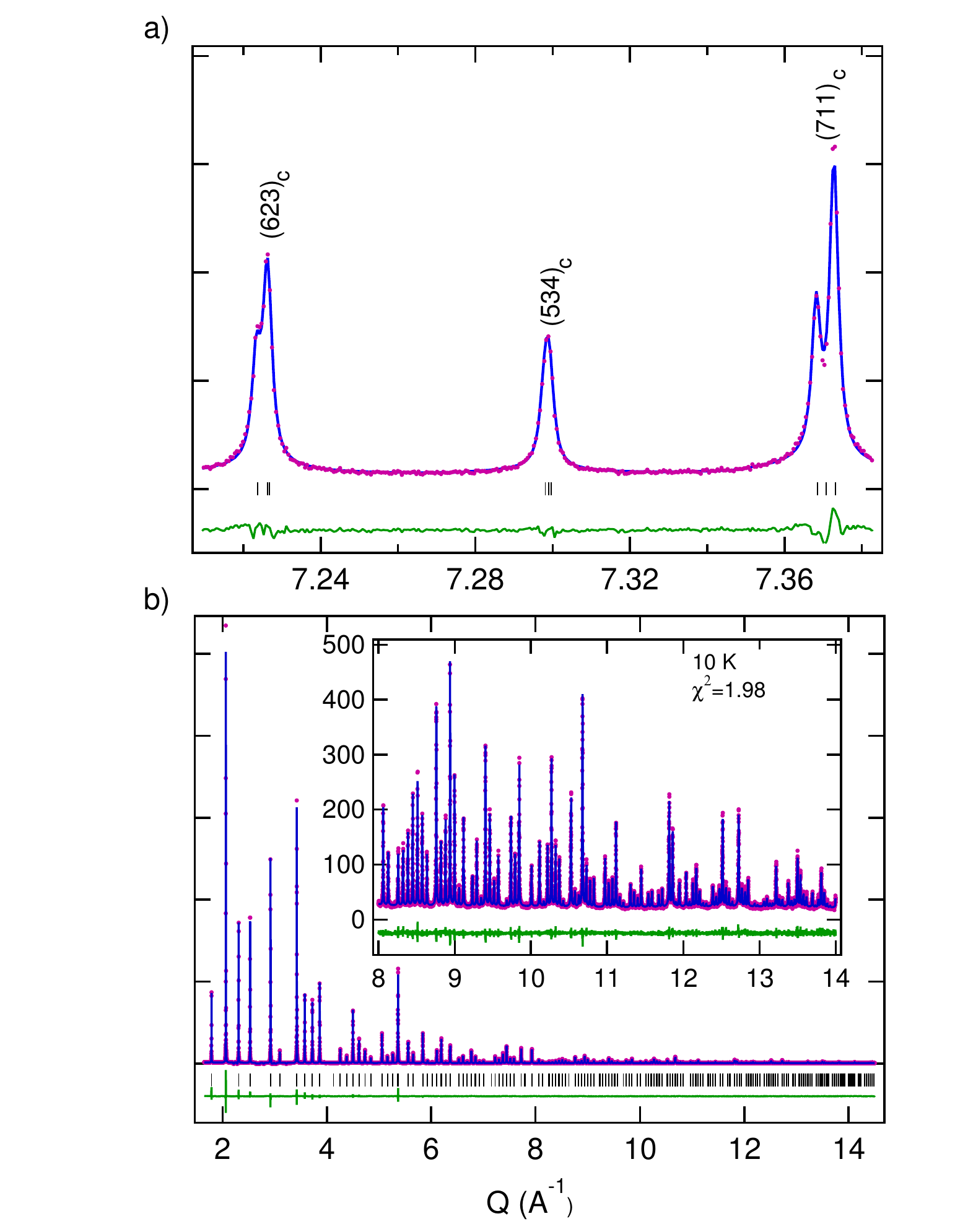}
\caption{(color online) (a) Observed, calculated and difference profiles for the Rietveld fit to a portion of the synchrotron x-ray diffraction profile of \mn~at 10 K. The pseudo-tetragonal peak splitting can be easily resolved. b) The complete range of fitted data, profiles and tick markers are as above.}
\label{Fig1}
\end{center}
\end{figure}

Turning now to the effect of low temperature on the crystal structure, the (511)$_{c}$ reflection is shown in the form of a colour map for the short scans collected on cooling from 55 - 10 K. A splitting can clearly be seen to emerge below $\sim$40 K. It is possible that the slight offset in temperature between our results and previous measurements are due to the rapid cooling employed.  Note that these data sets were binned with a step size of 0.003, $^{\circ}$ (comparable to the instrumental resolution) and it is only due to the high degree of crystallinity that the structural phase transition is observed at all. From these rapid scans, the distortion appears to be tetragonal.\\
The data set collected at 10 K has a significantly higher information content, extending out to Q=14.5 \AA$^{-1}$, with excellent statistics in the entire range (Fig. 5b). We therefore compared a range of models to the data, using both the Le Bail approach to confirm the lattice symmetry, and Rietveld refinements of several candidate structures. Before giving details of the main results, we note that the parent phase space group $Pa\bar{3}$~is one of the few cubic space groups which does not contain four-fold rotations. Although the (likely) first-order nature of the structural transition means that group-subgroup relations need not apply, this means that there are no ordered structures in simple tetragonally distorted cells, unless the disulphide groups lie along the direction of the  four-fold rotation axis. Such  a wholesale rotation of the disulphide groups seems less likely \cite{raman}~in \mn, compared to the true molecular crystal pyrites like the alkali metal superoxides \cite{NaO2}. As there is also no evidence for superstructure reflections resulting from an increase in unit cell size, we therefore explored the possibility of a pseudo-tetragonal orthorhombic structure. The space group $Pbca$~is a sub-group of $Pa\bar{3}$, and allows the construction of a model without generating new positions. Freely refining the unit cell parameters always resulted in cells which were metrically tetragonal within error. Note that, using the same double Lorentzian peak shape as before, there was no need to assume anisotropic peak shape broadening. The $a$~and $b$~axes were therefore constrained to be equal to reduce correlations in the final fit. The same constraints were added for the sulphur $x$~and $y$~coordinates. The final unit cell parameters were $a$~=$b$=6.085660(4) and $c$=6.089694(4) \AA. Full details of the refined atomic coordinates for the refinements at both 10 and 55 K can be found in Table I.
\begin{table}
\caption{\label{tab:table1}Refined lattice parameters and atomic coordinates for \mn~at 10 K ($Pbca$) and 55 K ($Pa-3$) from synchrotron x-ray powder diffraction. Note that the Wyckoff positions are Mn : $4a$ (0,0,0) and S: $8c$, in both phases.}
\begin{ruledtabular}
\begin{tabular}{cccc}
  & 55 K &10 K &\\
\hline
$a$~(\AA)&6.0888214(5)&6.085660(4)\\
$b$~(\AA)&6.0888214(5)&6.085660(4)\\
$c$~(\AA)&6.0888214(5)&6.089694(4)\\
volume~(\AA$^{3}$)& 225.7354(5) &225.533(6) \\
S: $x$&0.40006(1) &0.4006(4)\\
S: $y$& &0.4006(4)\\
S: $z$& &0.4008(4)\\
    \end{tabular}
\end{ruledtabular}
\end{table}\\

As the Mn cation remain on special positions, the structural distortion in \mn~has only a very weak effect on the bond distances. Naively, this would be expected to only weakly perturb the magnetic exchange interactions responsible for the long-range magnetic order. However, reviewing the available evidence \cite{MnS2_TC3,MnS2_TC4}, shows that the magnetic ground state of \mn~is remarkably anisotropic for an essentially cubic material. For example, the correlation length of the diffuse magnetic scattering observed well above T$_{N}$ is significantly longer \cite{MnS2_TC3,MnS2_TC4}~parallel to the (incommensurate) propagation vector. Deep within the ordered state, preliminary inelastic neutron measurements also detect two excitation branches, one of which is almost dispersion-less \cite{MnS2_TC4}. We speculate, that given the layered nature of the proposed type-III magnetic order, that the magnetic coupling is anomalously weak in the stacking direction. This is likely to correspond to the $c$~axis in our crystal structure model, and with our new result it should now be possible to understand the twinning which previously hampered low temperature investigations. Future investigations will therefore focus on the surprising emergence of anisotropy in this three-dimensional classical frustrated magnet. Finally, we note the existence of another tantalising report \cite{konrad}~of symmetry breaking in the $fcc$~magnet K$_{2}$IrCl$_{6}$. This materials shows a very similar weak lattice distortion just below T$_{N}$=3.1 K. To our knowledge, this is the only other $fcc$~material showing type-III magnetic order, raising the possibility that the behaviour reported here is universal in this class of magnets. \\

\noindent{\large{\textbf{Conclusions}}}\\
In summary, we have reported the characterisation of a new polymorph of \mn~at low temperature. This phase is pseudo-tetragonal, although symmetry arguments lead us to propose the orthorhombic space group $Pbca$. Following the suggestion of McCrone \cite{mccrone}, polymorphs should be labelled using roman numerals in order of discovery. Although we previously (2014) reported the structure of a new monoclinic polymorph found above 11.7 GPa, some evidence for the low temperature distortion discussed here was presented by Chattopadhyay and Liss in 1997. The polymorphs of \mn~ should thus be reported as \mn-I ($Pa\bar{3}$, 40 - 533 K, ambient pressure), \mn-II ($Pbca$, $<$ 40 K, ambient pressure) and \mn-III (\textit{P2$_{1}$/c,}$>$ 11.7 GPa, ambient temperature). We believe that the diversity found in this simple binary mineral points to strong coupling between spin and lattice degrees of freedom. In the present work, this is likely driven by geometrical frustration of the Heisenberg moments associated with the Mn$^{2+}$~cations. The orthorhombic distortion discovered here effectively lifts frustration, setting the stage for the establishment of long range magnetic order.\\

\noindent{\large{\textbf{Acknowledgements}}}\\
{\normalsize We thank Jon Wright for useful discussions, and acknowledge the ESRF for access to instrumentation and A.H. Hill for experimental assistance.}

\end{document}